\documentclass[%
reprint,
superscriptaddress,
showkeys,
amsmath,amssymb,
aps,
twocolumns]{revtex4-2}

\usepackage{amssymb}
\usepackage{amsmath}
\usepackage{amsthm}
\usepackage{natbib}
\usepackage{eucal}
\usepackage{mathrsfs}
\usepackage{graphicx}
\usepackage{dcolumn}
\usepackage{color}
\usepackage{bm}
\usepackage{hyperref}
\usepackage[usenames,dvipsnames,x11names,table]{xcolor}
\usepackage{balance}
\usepackage{dsfont}
\usepackage{tikz}
\usetikzlibrary{shapes}
\usepackage{multirow}
\usepackage{tabularx}
\usepackage{enumitem}
\usepackage{cancel,soul}

\setlist[itemize]{align=parleft,left=0pt..1em}
\newcommand{\Ra}{\operatorname{Ra}}
\newcommand{\Nu}{\operatorname{Nu}}
\newcommand{\We}{\operatorname{We}}
\newcommand{\I}{\operatorname{I}}

\makeatletter
\let\@internalcite\cite
\def\cite{\def\citeauthoryear##1##2{##1, ##2}\@internalcite}
\def\shortcite{\def\citeauthoryear##1##2{##2}\@internalcite}
\def\@biblabel#1{\def\citeauthoryear##1##2{##1, ##2}[#1]\hfill}
\makeatother
\begin{document}
\preprint{APS/123-QED}
\title{Role of interfacial stabilization in the Rayleigh-Bénard convection of liquid-liquid dispersions}
\author{Francesca Pelusi}
\email{francesca.pelusi@cnr.it}
\affiliation{Istituto per le Applicazioni del Calcolo, CNR - Via Pietro Castellino 111, 80131 Naples, Italy}
\author{Andrea Scagliarini}
\affiliation{Istituto per le Applicazioni del Calcolo, CNR - Via dei Taurini 19, 00185 Rome, Italy}
\affiliation{INFN, Sezione Roma ``Tor Vergata", Via della Ricerca Scientifica 1, 00133 Rome, Italy}
\author{Mauro Sbragaglia}
\affiliation{Department of Physics \& INFN, Tor Vergata University of Rome, Via della Ricerca Scientifica 1, 00133 Rome, Italy}
\author{Massimo Bernaschi}
\affiliation{Istituto per le Applicazioni del Calcolo, CNR - Via dei Taurini 19, 00185 Rome, Italy}
\author{Roberto Benzi}
\affiliation{Sino-Europe Complex Science Center, School of Mathematics \\North University
of China, Shanxi, Taiyuan 030051, China}
\affiliation{Department of Physics \& INFN, Tor Vergata University of Rome, Via della Ricerca Scientifica 1, 00133 Rome, Italy}
\date{\today}
\begin{abstract}
\noindent
Based on mesoscale lattice Boltzmann numerical simulations, we characterize the Rayleigh-Bénard (RB) convective dynamics of dispersions of liquid droplets in another liquid phase. Our numerical methodology allows us to modify the droplets' interfacial properties to mimic the presence of an emulsifier (e.g., a surfactant), resulting in a positive disjoining pressure that stabilizes the droplets against coalescence. To appreciate the effects of this interfacial stabilization on the RB convective dynamics, we carry out a comparative study between a proper emulsion, i.e., a system where the stabilization mechanism is present ({\it stabilized} liquid-liquid dispersion), and a system where the stabilization mechanism is absent ({\it non-stabilized} liquid-liquid dispersion). The study is conducted by systematically changing both the volume fraction, $\phi$, and the Rayleigh number, $\Ra$. We find that the morphology of the two systems is dramatically different due to the different interfacial properties. However, the two systems exhibit similar global heat transfer properties, expressed via the Nusselt number $\Nu$. Significant differences in heat transfer emerge at smaller scales, which we analyze via the Nusselt number defined at mesoscales, $\Nu_{\mathrm{mes}}$. In particular, stabilized systems exhibit more intense mesoscale heat flux fluctuations due to the persistence of fluid velocity fluctuations down to small scales, which are instead dissipated in the interfacial dynamics of non-stabilized dispersions. 
For fixed $\Ra$, the difference in mesoscale heat flux fluctuations depends non-trivially on $\phi$, featuring a maximum in the range $0.1 < \phi < 0.2$. Taken all together, our results highlight the role of interfacial physics in mesoscale convective heat transfer of complex fluids.
\end{abstract}
\keywords{Emulsions, multiphase flows, Rayleigh-B\'enard thermal convection}   

\maketitle

\section{Introduction}\label{sec:intro}
Emulsions are dispersions of liquid droplets in another liquid phase, stabilized by a surfactant that inhibits droplet coalescence~\cite{ravera2021,Tcholakova2004,Barnes94,Mason99,Derkach09,Tadros13}, thus resulting in a non-trivial collective response of the system when solicited with an external forcing~\cite{Pal2000,Bonn17}.
Emulsions are encountered in a plethora of settings, ranging from food~\cite{Mcclements15} and pharmaceutical industrial processes~\cite{Khan11}, to geophysics~\cite{Davaille2018} and energy technology~\cite{Goodarzi19}. 
However, their behavior in thermal flows is still poorly understood. The study of thermally driven emulsions -- and more generally thermally driven soft materials -- finds important applications in geophysics to simulate, for instance,  mantle convection from the interior to the planet surface~\cite{Orowan65,Morgan71,Montelli06,French15,Davaille2018}; in petrochemical industry~\cite{MartiMestres02}, where this kind of materials are central to enhanced oil recovery, separation, and refining operations and where temperature-dependent interfacial behavior and rheology affect flow, phase stability, and energy efficiency~\cite{abdulredha2020overview}; in food industry~\cite{Bai21}, where stability and nutrient maintenance may be altered by heat transfer induced in emulsion-based aliments. Recently, emulsions have found applications as working fluids in heat exchangers, especially in combination with microencapsulated phase change materials~\cite{delgado2012,Wang2019review}.\\
From the point of view of fundamental fluid mechanics the problem is outstanding: we are dealing with thermal flows and the associated instabilities~\cite{Benard1900,Rayleigh1916,Chandrasekhar61}, where velocity and thermal fluctuations are non trivially coupled and produce complex time-dependent flows featuring multiscale properties~\cite{Sun08,Sugiyama10,Ahlers09,Lohse10,Chilla12,Scheel13,Ecke23,Zhang2024,lohse2024ultimate}; the complexity of the flow is two-way coupled with a dispersion of deformable droplets. Depending on flow conditions, droplets can further undergo break-up and/or coalescence~\cite{Stone94,Cristini2003,Vankova07a,Vankova07b,Vankova07c,Perlekar2012,Scarbolo15,Roccon17,Soligo19,Liu21,Girotto22,Girotto24,CrialesiEsposito2024small,Brandt24,Mangani24,PelusiPRE25}, thus resulting in dynamical morphological changes in the emulsion. Due to this complexity, theoretical analyses, if any, exist only limited to simple systems~\cite{Zhang06,BalmforthRust09,Vikhansky09,Vikhansky10,AlbaalbakiKhayat11,Balmforthetal14,Karimfazli16,Aghighi23}. Moreover, experiments could face several limitations in exploring thermal convection in emulsions as a result of both the complex nature of these materials and technical limitations. On the one hand, controlling the stability and droplet size distribution over time is challenging, as emulsions are prone to coalescence~\cite{Plassard24}, phase inversion~\cite{Bouchama03,Perazzo15}, and Ostwald ripening~\cite{Weiss2000}, whereas, on the other hand, optical opacity limits flow visualization and measurement~\cite{Hu17}. As a result, numerical simulations~\cite{Urbina09,Aland12,Turanetal12,Massmeyer13,Gupta15hybrid,Hassanetal15,Karimfazli16,Rosti19numerical,PelusiSM21,Pelusi24rheology,TiribocchiReview25} often remain the most effective, and sometimes the only, means to investigate the intricate dynamics of multiphase and droplet-laden thermal flows. Indeed, various numerical studies were conducted to characterize the heat transfer properties in different kinds of dispersions, such as thermally bounded convective flows laden with droplets~\cite{PelusiSM21,PelusiSM23,Brandt24,Mangani24} or bubbles (the limiting situation of droplets with negligible internal viscosity)~\cite{Orestaetal09,Lakkaraju13,biferale12,biferale2013simulations}, thermally-driven multilayer configurations \cite{Liu21,Liu22}, also with different boundary conditions~\cite{Liu21a}. These studies revealed that dispersed objects (droplets/bubbles) can impact thermal plumes, heat transfer properties, and energy transfer at small scales. Other studies considered the thermal behavior of shear thinning/non-Newtonian~\cite{BalmforthRust09,AlbaalbakiKhayat11} or viscoplastic fluids with a finite yield-stress~\cite{Zhang06,Vikhansky09,Vikhansky10,Turanetal12,Massmeyer13,Balmforthetal14,Hassanetal15,Karimfazli16,santos21}, without explicitly considering the presence of dispersed droplets, but rather accounting for their presence via an effective remodulation of the constitutive relation between the shear-rate and the stress in the continuum equations. These studies highlighted the importance of the role that the non-Newtonian rheology -- and in particular a finite yield stress~\cite{Bonn17} -- has on the flow initiation and the subsequent thermal plumes dynamics. Still, these studies did not explicitly consider break-up and/or coalescence of droplets and/or their plasticity at mesoscales~\cite{Goyon08,Goyon10}, which have a non-trivial impact in thermally convective flows, as highlighted by experiments on yield-stress fluids~\cite{Davailleetal13,Hassanetal15} and recent numerical simulations by the authors~\cite{PelusiPRL24,PelusiPRE25}. In particular, the interplay between droplet break-up/coalescence and buoyancy forces gives rise to a variety of thermal convection regimes, depending on the droplet concentration $\phi$ and intensity of buoyancy~\cite{PelusiPRE25}: for larger values of $\phi$, coalescence phenomena dominate dynamics, eventually leading to phase inverted states exhibiting an intermittent transient dynamics~\cite{PelusiPRL24}; for sufficiently large buoyancy forces, phase inversion inevitably accompanies a sustained convection; finally,  when $\phi$ is small, break-up phenomena dominates dynamics.\\
An essential ingredient in the preparation of emulsions is the presence of an emulsifier (i.e., surfactant), which produces a positive disjoining pressure inside the thin liquid film between the interfaces of two adjacent droplets~\cite{Chen05experimental,Nesterenko14,ravera2021}. This disjoining pressure results in an interfacial stabilization mechanism that prevents the coalescence of droplets, leading to the formation of emulsions as stabilized liquid-liquid dispersions~\cite{Tcholakova2004,Barnes94,Mason99,Derkach09,Tadros13}. This mechanism is crucially influential in the physics of emulsions, especially under concentrated conditions, where droplet-droplet contacts are favored and a closely packed droplet configuration would otherwise be unfeasible. Moreover, the capability of storing elastic energy among neighbouring deformable droplets~\cite{BarratReview17,Nicolas18,Divoux24}, endows the emulsion with non-Newtonian rheological properties~\cite{Bonn17}. This feature leads to a {\it distinctive} difference, both morphologically and rheologically, between stabilized and non-stabilized liquid-liquid dispersions. The latter, in particular, cannot sustain high droplet volume fraction and, therefore, are inherently Newtonian. In the setting of convective thermal flows, it is then natural to expect different dynamical behaviors, a fact that has not been systematically highlighted in the literature so far. Our paper aims to fill this gap. Using hydrodynamic numerical simulations based on the lattice Boltzmann method (LBM) and working in the paradigmatic setup of Rayleigh-Bénard (RB) convection~\cite{Benard1900,Rayleigh1916,Chandrasekhar61,Grossmann01,Ahlers09,Lohse10,Chilla12}, we simulate the buoyancy-driven dynamics of a stabilized liquid-liquid dispersion (i.e., a proper emulsion), as discussed in previous studies~\cite{PelusiSM21,PelusiSM23,PelusiPRL24,PelusiPRE25}, and we compare it with the dynamics of a non-stabilized liquid-liquid dispersion, missing a positive disjoining pressure. We explore the various emerging dynamical regimes by changing the volume fraction of dispersed phase, $\phi$, and the buoyancy force, quantified in terms of the dimensionless Rayleigh number, $\Ra$. We analyze both the heat flux and the system morphology (via the use of interface indicators) at changing $\phi$ and $\Ra$. We find that the morphologies of the two systems significantly differ; however, the dependence of the {\it global} heat flux properties on $\phi$ and $\Ra$ does not seem sensitive to the physicochemical characteristics of the dispersion (namely, whether it is stabilized or not). A {\it mesoscale} analysis of the heat flux properties reveals, instead, a mismatch between the two systems, featuring enhanced fluctuations in the stabilized system for all volume fractions $\phi <  0.5$, with a maximum relative increase at a ($\Ra$-dependent) volume fraction in the range $0.1 < \phi < 0.2$. The presence of a maximum is arguably related to the coupling between small-scale velocity fluctuations and interface dynamics, which is markedly impacted by the interfacial stabilization.\\

The paper is organized as follows: in Sec.~\ref{sec:method}, we describe the system setup, introduce the dimensionless control parameters, and provide a summary of the LBM employed. Results of numerical simulations are shown in Sec.~\ref{sec:results}. Finally, conclusions are given in Sec.~\ref{sec:conclusions}.

\section{Model and simulations}\label{sec:method}
We first describe the hydrodynamic setup and associated dimensionless numbers considered in the numerical simulations in Sec.~\ref{sec:setup}; we then review the numerical methodology used in Sec.~\ref{sec:LBM}.
\begin{figure*}[t!]
    \centering    \includegraphics[width=.9\textwidth]{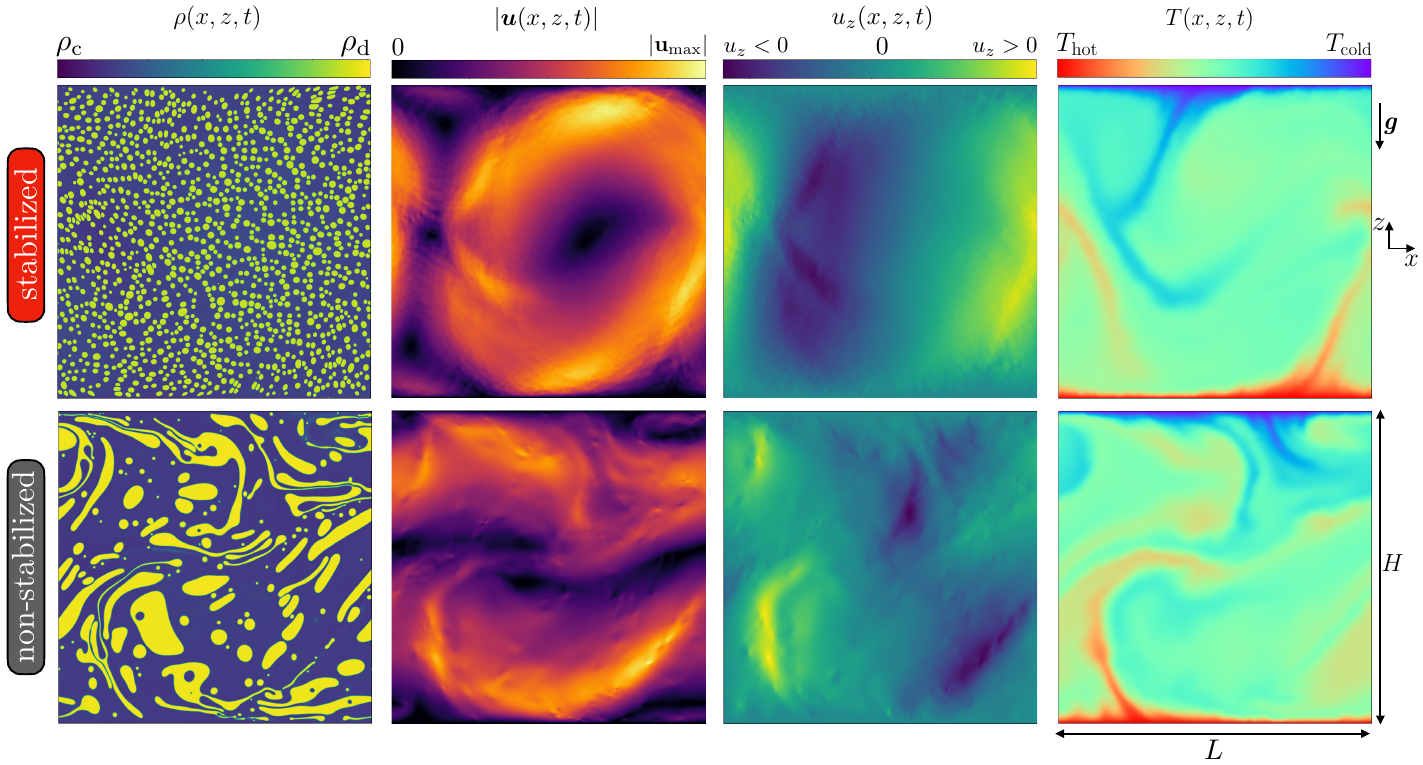}
    \caption{Representative snapshots of the dynamics of the Rayleigh-Bénard (RB) thermal convection of stabilized and non-stabilized liquid-liquid dispersions. Systems are confined between a lower hot and an upper cold wall at a distance $H$, thus undergoing a temperature difference $\Delta T = T_{\mathrm{hot}} - T_{\mathrm{cold}}$. A buoyancy force is applied due to the gravity ${\bm g}$. Convective states are analyzed by investigating the hydrodynamical fields of density $\rho (x,z,t)$, velocity ${\bm u}(x,z,t)$, and temperature $T (x,z,t)$. Snapshots refer to systems with a volume fraction $\phi=0.28$ and Rayleigh number $\Ra \approx 1.6 \times 10^{7}$.}
    \label{fig:sketch_setup}
\end{figure*}
\subsection{Hydrodynamical setup and dimensionless parameters}\label{sec:setup}
For both stabilized and non-stabilized liquid-liquid dispersions, we consider a collection of dispersed (d) liquid droplets with kinematic viscosity $\nu$ in another equiviscous carrier (c) liquid in a two-dimensional setup of RB thermal convection~\cite{Grossmann01,Ahlers09,Chilla12} with coordinates $(x,z)$, wherein two parallel plates of length $L$ extending in the $x$ direction are placed at distance $H$ and located at $z=\pm H/2$ (see Fig.~\ref{fig:sketch_setup}). In contrast to previous works~\cite{PelusiSM21,PelusiSM23,PelusiPRL24,PelusiPRE25}, where simulations were performed with an aspect ratio $\Gamma = L/H = 2$, we adopted a fixed aspect ratio of $\Gamma = 1$ throughout this study (that is, $H=L$). This choice is motivated by the phenomenology observed in the non-stabilized liquid-liquid dispersion at low values of $\Ra$, where, for a smaller cell height (i.e. stronger confinement), it is observed that a large droplet of dispersed phase can "squeeze" between the walls before breaking in daughter droplets that remain trapped within the convective rolls.  By doubling the cell height at fixed width, i.e., working at $\Gamma=1$ (weaker confinement), we observe that such an effect is mitigated. This phenomenology surely warrants detailed studies in a separate work. The system is heated from below and cooled from above, fixing at all times ($t$) the bottom and upper plate temperatures at $T(x,z=-H/2,t)=T_{\text{hot}}$ and $T(x,z=+H/2,t)=T_{\text{cold}}$, resulting in a temperature jump $\Delta T=T_{\text{hot}}-T_{\text{cold}}$; periodic boundary conditions are applied in the $x$ direction (see Fig.~\ref{fig:sketch_setup}). \\
The dispersed and carrier phases are described by density fields $\rho_{\text{d,c}}(x,z,t)$; as detailed below, the methodology used for the numerical simulations implies diffuse interfaces, i.e., an interface with small (but finite) thickness that separates regions with the majority of the dispersed phase ($\rho_{\text{d}} \gg \rho_{\text{c}}$) from regions with the majority of the carrier phase ($\rho_{\text{d}} \ll \rho_{\text{c}}$). Further, the presence of interfaces is associated with a surface tension $\Sigma$. The initial configuration consists of $N_{\text{d}}$ droplets with a predefined spatial arrangement, which is generated through a preparatory simulation~\cite{TLBfind22,PelusiPRE25}: the centers-of-mass of the droplets are initially positioned in a honeycomb-like pattern, and we assign to the droplets a circular geometry of diameter $d$. Then, to introduce disorder and facilitate a physically realistic setup, small random perturbations are applied to both the droplets' initial centers-of-mass positions and the density field of the surrounding continuous liquid. The system is then allowed to evolve freely with no applied forcing, relaxing towards a lower-energy configuration, thus reaching a slightly polydisperse droplet size distribution. The resulting system is then identified with the volume fraction 
\begin{equation}\label{eq:phi}
\phi=\frac{A_{\text{d}}}{L H} \ ,
\end{equation}
where ${A_{\text{d}}}$ is the total area occupied by the dispersed phase. In the presence of diffuse interfaces, we need to introduce a threshold to compute $A_{\text{d}}$ as $A_{\text{d}} = \int \int \Theta(\rho_{\text{d}}(x,z)-\rho^*)dx \ dz$, where $\Theta$ is the Heaviside step function and $\rho^*$ is a threshold value corresponding to the mean value between the dispersed and carrier phase in the bulk region~\cite{PelusiSM21,PelusiSM23}. We distinguish between stabilized and non-stabilized liquid-liquid dispersions by introducing a mechanism that promotes, in the former case, the emergence of a positive disjoining pressure in the thin liquid films separating closely adjacent interfaces, which inhibits droplet coalescence. Then, gravity acts in the negative $z$ direction with strength $g$, resulting in a buoyancy force density equal to $\rho \alpha (T-T_0) g$, where $T_0$ is a reference temperature~\footnote{For all simulations, we fix $T_0 = 0$.}, $\rho=\rho_\mathrm{d}+\rho_\mathrm{c}$ is the total density, and $\alpha$ is the thermal expansion coefficient. Once the droplets are placed in the desired initial condition with a given volume fraction $\phi$, we apply a suitable sinusoidal perturbation to the velocity field to trigger convection~\cite{PelusiPRL24}. Convective states are analyzed via the hydrodynamical velocity, ${\bm u}(x,z,t)$, the density of the dispersed and carrier phase, $\rho_{\text{d,c}}(x,z,t)$, and the temperature field $T(x,z,t)$ (see Fig.~\ref{fig:sketch_setup}). The dynamics of the temperature field $T(x,z,t)$ is governed by an advection-diffusion equation, with the advection set by the hydrodynamical velocity and the diffusion regulated by a constant thermal diffusivity $\kappa$. In all numerical simulations, we use, as input parameters, the volume fraction $\phi$ defined in Eq.~\eqref{eq:phi} and the Rayleigh number,~\cite{Chandrasekhar61,Bodenschatz2000,Ahlers09,Lohse10,Chilla12}
\begin{equation}\label{eq:Ra}
\Ra = \frac{\alpha g \Delta T H^3}{\nu \kappa} \ ,
\end{equation}
which encodes the buoyancy intensity. The Prandtl number Pr$=\nu/\kappa$ is kept fixed in all simulations. Notice that, in contrast to homogeneous fluids, stabilized liquid-liquid dispersions exhibit a response to external forcing that is significantly influenced by interfacial effects, the latter altering the viscous, elastic, and visco-plastic properties of the system. Consequently, $\Ra$ in Eq.~\eqref{eq:Ra} loses its dynamical meaning for the stabilized case as the volume fraction $\phi$ increases, since the effective viscosity of the system becomes sensitive to the local shear rate. Therefore, throughout this work, we define both $\Ra$ and $\Pr$ considering the kinematic viscosity of the corresponding homogeneous reference fluid with $\rho_{\mathrm{d}} = \rho_{\mathrm{c}}$. Under this convention, $\Ra$ remains a dimensionless parameter that quantifies the intensity of the applied buoyancy. We assume no variation in surface tension $\Sigma$ between the two systems under comparison (see Sect.~\ref{sec:LBM} for details). As a consequence, we do not consider the Weber number, $\We = \rho U^2 d / \Sigma$ ($U$ is the characteristic flow velocity), as an independent control parameter, because $\Ra$ directly determines it in this context~\footnote{For the values of $\Ra$ explored in this work, the Weber number is in the range $\We \in [0.1,40]$}.\\
To monitor the heat transfer properties, in every point ${\bm x}=(x,z)$ and for any time $t$ we compute the corresponding local Nusselt number as
\begin{equation}\label{eq:Nu_space-time}
\Nu({\bm x},t) = 1 + \frac{u_z({\bm x},t) T({\bm x},t)} {\kappa \frac{\Delta T}{H}}\ ,
\end{equation}
and we average in space to obtain the global Nusselt number $\Nu$~\cite{Shraiman90,Ahlers09,Verzicco10,Chilla12}
\begin{equation}\label{eq:Nusselt}
\Nu(t) = \langle \Nu({\bm x},t)\rangle_{x,z} \ ,
\end{equation}
where $\langle \dots \rangle_{x,z}$ refers to the spatial average on the overall domain $L \times H$. In the statistically steady state (hereafter referred to as ``steady-state'', for brevity), we average the signal $\Nu(t)$ over time, thus computing $\overline{\Nu} = \langle \Nu \rangle_t$. The system morphology is assessed in terms of the interface indicator defined as:
\begin{equation}\label{eq:II}
    \I(t) = \langle |\nabla \chi ({\bm x},t)| \rangle_{x,z}, \quad \chi ({\bm x},t) = \frac{\rho_d({\bm x},t) - \rho_c({\bm x},t)}{\rho_d({\bm x},t) + \rho_c({\bm x},t)},
\end{equation}
where $\chi$ is the phase field. Similarly to the study of heat transfer, we average $\I(t)$ over time when focusing on steady-state properties, obtaining $\overline{\I} = \langle \I \rangle_t$.
\subsection{Lattice Boltzmann model}\label{sec:LBM}
Simulations are carried out using the lattice Boltzmann methods (LBMs)~\cite{Benzi92,Kruger17,Succi18}, which are computational tools capable of accurately capturing the behavior of multiphase and multicomponent systems~\cite{Benzietal14,Dollet15,Leclaire17,LulliBenziSbragaglia18,PelusiSM19,PelusiEPL19,Girotto24,PelusiSM21,PelusiPOF22,Pelusi24rheology,PelusiPRL24,TiribocchiReview25}. Particularly, we employ the open-source code \texttt{TLBfind}~\cite{TLBfind22}, featuring the dynamics of a multicomponent system in thermal flows. The essential aspects are summarized below, while a more comprehensive treatment is presented in Ref.~\cite{TLBfind22}. Hereafter, all dimensional quantities are given in lattice Boltzmann simulation units.\\
The system dynamics is simulated by evolving on a two-dimensional lattice the distribution function $f_{\xi,i}({\bm x},t)$, which represents the probability of finding a fluid particle of phase $\xi = \mathrm{d,c}$ with discrete velocity ${\bm c}_i$~\footnote{The model employs a two dimensional D2Q9 lattice with 9 velocities ${\bm c}_i$ indexed from $i=0, \dots , 8$: ${\bm c}_0 = (0,0),\quad {\bm c}_{1,3} = (\pm1,0),\quad {\bm c}_{2,4} = (0,\pm1),\quad {\bm c}_{5,8} = (+1,\pm1),\quad {\bm c}_{6,7} = (-1,\pm1)$.} at location ${\bm x} = (x,z)$ and time $t$. Spatial and temporal spacings are set to $\Delta x = \Delta t = 1$, respectively. 
The dynamical evolution of $f_{\xi,i}$ is governed by the lattice Boltzmann equation
\begin{equation}\label{eq:LBM}
f_{\xi,i}({\bm x}+{\bm c}_i,t+1) -  f_{\xi,i}({\bm x},t) = -\frac{1}{\tau}\left[f_{\xi,i}({\bm x},t)-f_{\xi,i}^{(\mathrm{eq})}({\bm x},t)  \right] \ , 
\end{equation}
where $f_{\xi,i}^{(\mathrm{eq})}$ is a local equilibrium distribution function, $f_{\xi,i}^{(\mathrm{eq})}({\bm x},t)=f_{\xi,i}^{(\mathrm{eq})}(\rho_{\xi}({\bm x},t),\bar{\bf u}_\xi({\bm x},t))$, 
whose dependence from the density $\rho_{\xi}$ and  velocity $\bar{\bf u}_\xi$ (defined below) is set by
\begin{equation}\label{eq:feq}
f_{\xi,i}^{(\mathrm{eq})}(\rho_{\xi},\bar{\bf u}_\xi)= w_i \rho_{\xi} \left[1+\frac{\bar{\bf u}_\xi \cdot {\bm c}_{i}}{c_s^2}-\frac{\bar{\bf u}_\xi \cdot \bar{\bf u}_\xi}{2 c_s^2} + \frac{(\bar{\bf u}_\xi \cdot {\bm c_{i}})^2}{2 c_s^4} \right].
\end{equation}
The distribution function $f_{\xi,i}$ relaxes towards the local equilibrium $f_{\xi,i}^{(\mathrm{eq})}$ with a relaxation time $\tau$~\footnote{The relaxation time is fixed to $\tau = 1$ in all simulations.}. In Eq.~\eqref{eq:feq}, $c_s = 1/\sqrt{3}$ is the lattice speed of sound and $w_i$ are model-dependent weights~\cite{Kruger17,Succi18,TLBfind22}~\footnote{Weights $w_i$ in the D2Q9 scheme: $w_0 = 4/9$, $w_{1,4} = 1/9$, $w_{5,8} = 1/36$.}. The macroscopic density and velocity fields can be computed directly from $f_{\xi,i}$ as
\begin{equation}
\begin{split}
\rho_{\xi}({\bm x},t) &= \sum_{i=0}^{8} f_{\xi,i}({\bm x},t), \\
{\bf u}({\bm x},t) &= \frac{1}{\rho({\bm x},t)} \sum_{\xi} \sum_{i=0}^{8} {\bm c}_i f_{\xi,i}({\bm x},t),
\end{split}
\end{equation}
with $\rho = \sum_\xi \rho_\xi$ being the total density. Initial densities of the carrier and dispersed liquids are set to $\rho^{\mathrm{max}}_\mathrm{d,c} = 1.18$ and $\rho^{\mathrm{min}}_\mathrm{d,c} = 0.18$, respectively. Working in the framework of Shan-Chen pseudopotential LBMs~\cite{ShanChen93,Sbragaglia07,Benzi09,Sbragagliaetal12}, external and internal forces ${\bm F}_\xi$ acting on each component are incorporated in the dynamics (Eq.~\eqref{eq:LBM}) via the equilibrium velocity $\bar{{\bf u}}_\xi$ entering in Eq.~\eqref{eq:feq}, as
\begin{equation}\label{eq:ueq}
    \bar{{\bf u}}_\xi({\bm x},t) = {\bf u}({\bm x},t) + \frac{\tau {\bm F}_\xi({\bm x},t)}{\rho_\xi({\bm x},t)}.
\end{equation}
\begin{figure*}[t!]
    \centering    \includegraphics[width=.8\textwidth]{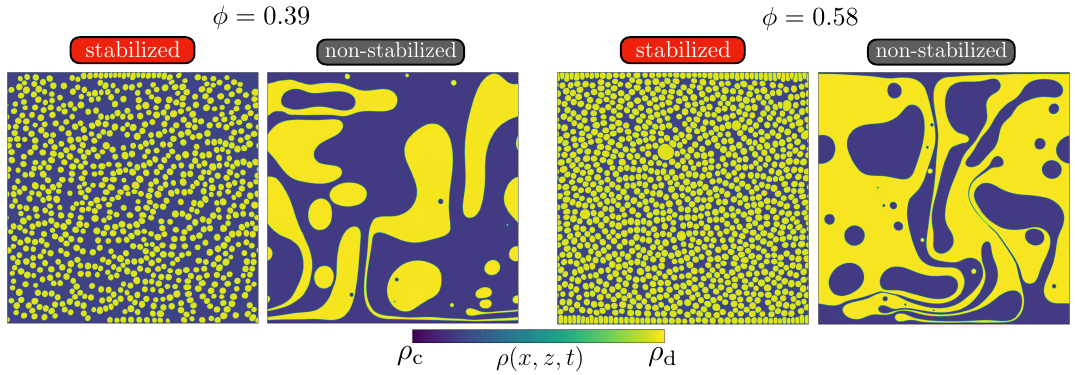}
    \caption{Comparison between stabilized and non-stabilized liquid-liquid dispersions at varying volume fractions $\phi$. The selected values of $\phi$ are chosen in such a way that $|\phi - \bar{\phi}| \approx 0.1$, with $\bar{\phi}=0.5$. While non-stabilized systems exhibit morphological symmetry under the exchange of continuous and dispersed phases, stabilized liquid-liquid dispersions break this symmetry due to the presence of a disjoining pressure. All snapshots correspond to simulations at $\Ra \approx 1.6 \times 10^{6}$.}\label{fig:comparison_phi_0.4_0.6}
\end{figure*}
Notice that the LBM framework reproduces the Navier-Stokes equations for the hydrodynamic velocity ${\bm u} = {\bf u} + {\bm F} / 2\rho$, with a kinematic viscosity $\nu = c_s^2 (\tau - 1/2)$~\cite{Kruger17,Succi18}. In Eq.~\eqref{eq:ueq}, ${\bm F}_\xi$ encompasses several contributions to the interaction as
\begin{equation}\label{eq:total_force}
{\bm F}_{\xi}({\bm x},t) ={\bm F}^{\mathrm{intra}}_{\xi}({\bm x},t)+{\bm F}^{\mathrm{inter}}_{\xi}({\bm x},t)+{\bm F}_{\xi}^{\mathrm{ext}}({\bm x},t) \ . 
\end{equation}
Specifically, ${\bm F}^{\mathrm{intra}}_{\xi}$ encodes the intra-component interactions and its definition is based on the pseudo-potential $\psi$:
\begin{equation}\label{eq:F12}
{\bm F}^{\mathrm{intra}}_\xi({\bm x},t) = -G_{\text{dc}} \psi_\xi({\bm x},t) \sum_{i=0}^{8} w_i \psi_{\xi'}({\bm x}+{\bm c}_i,t) {\bm c}_i,
\end{equation}
where $\xi' \neq \xi$, $\psi_{\xi,\xi'} = \rho_{\xi,\xi'} / \rho_0$ ($\rho_0$ is a reference density value~\footnote{We fix $\rho_0 = 0.83$.}). This interaction leads to phase separation and the formation of interfaces, thus the coupling parameter $G_{\text{dc}}$ primarily controls the surface tension $\Sigma$ at the interface. Furthermore, ${\bm F}^{\mathrm{inter}}_{\xi}$ includes both short-range attractive and mid-range repulsive forces within the same component
\begin{align}\label{eq:F_stabilization}
{\bm F}^{\mathrm{inter}}_\xi({\bm x},t) = & - G^a_{\xi\xi} \psi_\xi({\bm x},t) \sum_{i=0}^{8} w_i \psi_\xi({\bm x}+{\bm c}_i,t) {\bm c}_i \nonumber \\
& - G^r_{\xi\xi} \psi_\xi({\bm x},t) \sum_{i=0}^{24} p_i \psi_\xi({\bm x}+{\bm c}_i,t) {\bm c}_i,
\end{align}
where $G^a_{\xi \xi} < 0$, $G^r_{\xi \xi} > 0$ are interaction parameters, $p_i$ are weights over an extended lattice set~\cite{Benzi09,TLBfind22}, and the pseudo-potential $\psi_\xi$ follows the original Shan-Chen form $\psi_\xi = \rho_0 \left( 1 - \exp\left( -\rho_\xi/\rho_0 \right) \right)$~\cite{ShanChen93}.
Eq.~\eqref{eq:F_stabilization} constitutes the key feature of the model adopted in this work, as it effectively mimics the stabilizing role of surfactants at fluid interfaces by producing a positive disjoining pressure in the thin liquid films separating close-by interfaces ~\cite{Sbragagliaetal12,Dollet15}. As discussed earlier, this tends to prevent droplet coalescence, thereby enabling the formation of stabilized liquid-liquid dispersions. Stabilized liquid-liquid dispersions are obtained by setting the interaction parameters to $G^a_{\mathrm{cc}} = -9.0$, $G^a_{\mathrm{dd}} = -8.0$, $G^r_{\mathrm{cc}} = 8.1$, and $G^r_{\mathrm{dd}} = 7.1$, whereas they are set equal to zero in the non-stabilized case. It is also important to note that turning off the inter-component interactions (cfr. Eq.~\eqref{eq:F_stabilization}) alters the effective surface tension $\Sigma$. Hence, to ensure consistency between the two scenarios, i.e., to recover the exact value of $\Sigma \sim 0.0325$, we set $G_{\mathrm{dc}} = 0.405$ for the stabilized liquid-liquid dispersion. In contrast, the value was increased to $G_{\mathrm{dc}} = 0.65$ in the non-stabilized case. In other words, force contributions in Eq.~\eqref{eq:F_stabilization} allow to switch from a stabilized liquid-liquid dispersion to a non-stabilized one in the same computational framework.\\
\begin{figure*}[t!]
    \centering    \includegraphics[width=.85\textwidth]{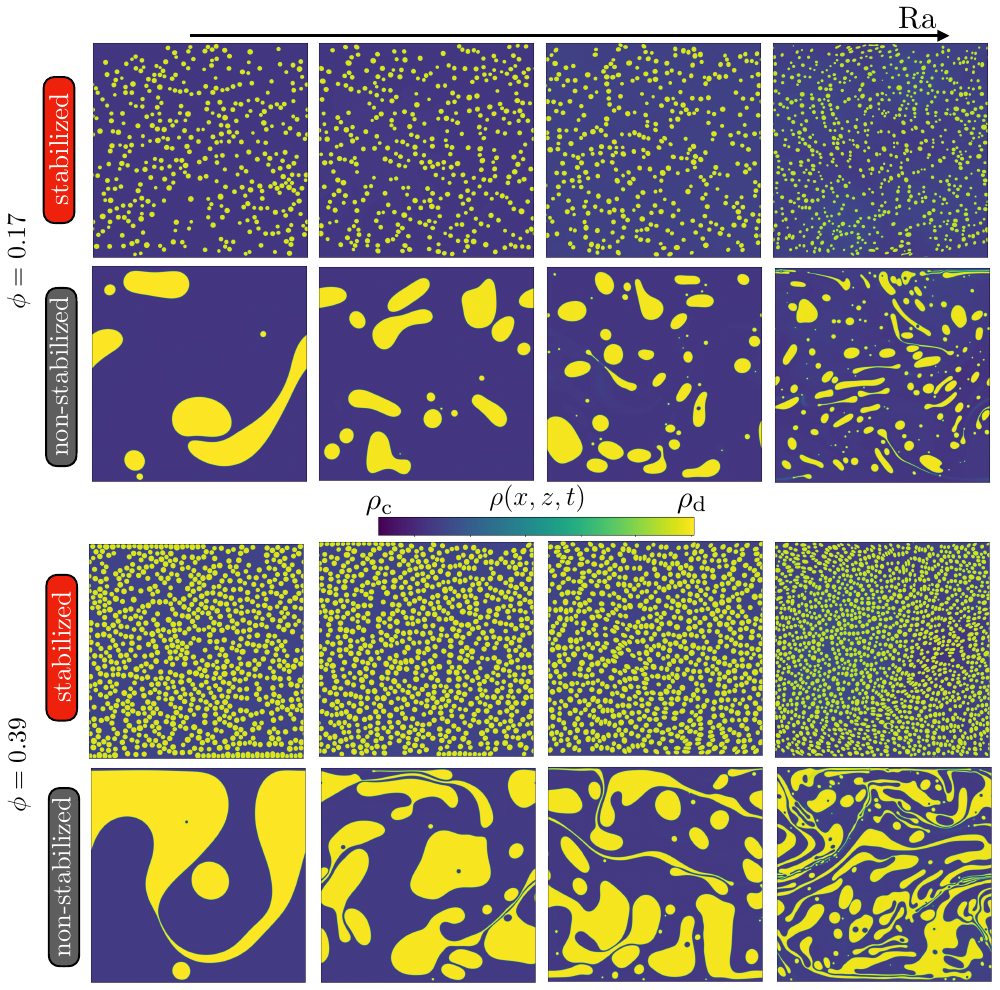}
    \caption{Density map snapshots for different volume fractions $\phi$ and different Rayleigh numbers $\Ra$, featuring both stabilized and non-stabilized liquid-liquid dispersions. $\Ra$ increases from left to right.}\label{fig:snapshots_Ra_phi}
\end{figure*}
The dynamics of the temperature field $T = T({\bm x},t)$ is incorporated into the model by adding a distribution function, $h_i({\bm x},t)$, which is governed by a dedicated lattice Boltzmann equation~\cite{Kruger17,Succi18}:
\begin{equation}\label{eq:LBM_thermal}
\small
h_{i}({\bm x}+{\bm c}_i,t+1) - h_{i}({\bm x},t) = -\frac{1}{\tau_h} \left[h_{i}({\bm x},t)-h_{i}^{(\mathrm{eq})}({\bm x},t) \right] \ ,
\end{equation}
where $\tau_h$ is the associated relaxation time and $h_{i}^{(\mathrm{eq})}$ is the local equilibrium distribution function, $h_{i}^{(\mathrm{eq})}({\bm x},t)=h_{i}^{(\mathrm{eq})}(T({\bm x},t),{\bm u}({\bm x},t))$, whose dependency from the temperature $T$ and velocity ${\bm u}$ is set by:
\begin{equation}\label{eq:teq}
h_{i}^{(\mathrm{eq})}(T,{\bm u})= w_i T \left[1+\frac{{\bm u} \cdot {\bm c}_{i}}{c_s^2}-\frac{{\bm u} \cdot {\bm u}}{2 c_s^2} + \frac{({\bm u} \cdot {\bm c_{i}})^2}{2 c_s^4} \right].
\end{equation}
Similarly to the fluid case, the macroscopic temperature $T({\bm x},t)$ is obtained as the zeroth-order moment of the distribution $h_i$, i.e.,
\begin{equation}
    T({\bm x},t)=\sum_{i=0}^{8} h_i({\bm x},t) .
\end{equation}
Thermal effects enter the dynamics of both fluid components through a buoyancy forcing term modeled via an external force ${\bm F}_{\xi}^{\mathrm{ext}}$ in Eq.~\eqref{eq:total_force} expressed in the Boussinesq approximation~\cite{Spiegel60}:
\begin{equation}\label{eq:F_ext}
{\bm F}_{\xi}^{\mathrm{ext}}({\bm x},t) = - \rho_\xi({\bm x},t) \alpha {\bm g} T({\bm x},t) .
\end{equation}
On the hydrodynamic scale, this formulation allows us to solve an advection-diffusion equation for $T$, where the temperature is advected by the fluid velocity ${\bm u}$ and the thermal diffusivity is given by $\kappa = c_s^2(\tau_h - 1/2)$.\\
Simulations are conducted on a $2048 \times 2048$ lattice grid. Droplets are initialized with a diameter $d \sim 50$ lattice spacings. Regarding boundary conditions, bounce-back rules are applied along the $z$ direction to both fluid components to enforce no-slip conditions at walls. At the same time, the temperature field is subject to Dirichlet conditions at the top and bottom boundaries~\cite{Kruger17,Succi18}. \\
We performed numerical simulations employing Nvidia A30 GPUs. In total, we carried out approximately 120 distinct numerical simulations, each typically requiring $\sim48$ GPU-hours. The simulation time, expressed in units of the characteristic free-fall time $t_{FF} \sim \sqrt{H/\alpha g \Delta T}$, varies according to the strength of the imposed buoyancy force: $t_{\mathrm{FF}}$ spans from roughly $6.4 \times 10^{3}$ (for $\Ra \approx 1.6 \times 10^7$) up to about $4.5 \times 10^{4}$ (for $\Ra \approx 3.5 \times 10^5$). Once the steady-state is attained, time-averaged quantities are computed over intervals ranging from 14 to 800 $t_{\mathrm{FF}}$, depending on $\Ra$. For the analysis of mesoscale observables, we gather statistics by including all droplets observed throughout the steady-state regime at all recorded time steps.
\section{Results and discussion}\label{sec:results}
\begin{figure*}[t!]
    \centering    \includegraphics[width=1.\textwidth]{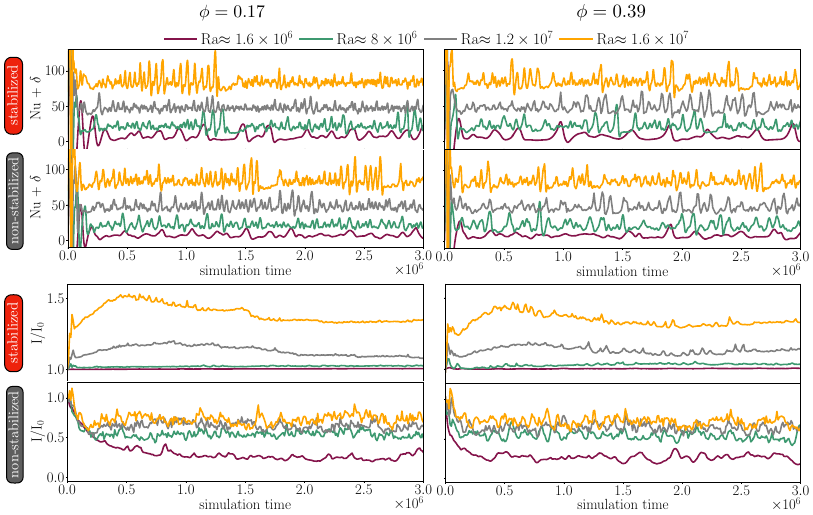}
    \caption{Time evolution of the Nusselt number $\Nu$ (cfr. Eq.~\eqref{eq:Nusselt}) and the interface indicator $\I$ (cfr. Eq.~\eqref{eq:II}), normalized with its initial value $\I_0$, for stabilized and non-stabilized liquid-liquid dispersions. Left panels refer to cases with a volume fraction $\phi=0.17$. Right panels refer to the case with $\phi=0.39$. Different values of the Rayleigh number $\Ra$ (different colors) are considered. To facilitate readability, data for $\Nu$ are vertically shifted by a quantity $\delta$ which depends on $\Ra$: $\delta = 0$ for $\Ra \approx  1.6 \times 10^{6}$, $\delta=10$ for $\Ra \approx 8 \times 10^{6}$, $\delta = 35$ for $\Ra \approx 1.2 \times 10^{7}$, while $\delta = 70$ for $\Ra \approx 1.6 \times 10^{7}$.}\label{fig:signals_vs_Ra}
\end{figure*}
\begin{figure*}[t!]
    \centering    \includegraphics[width=.8\textwidth]{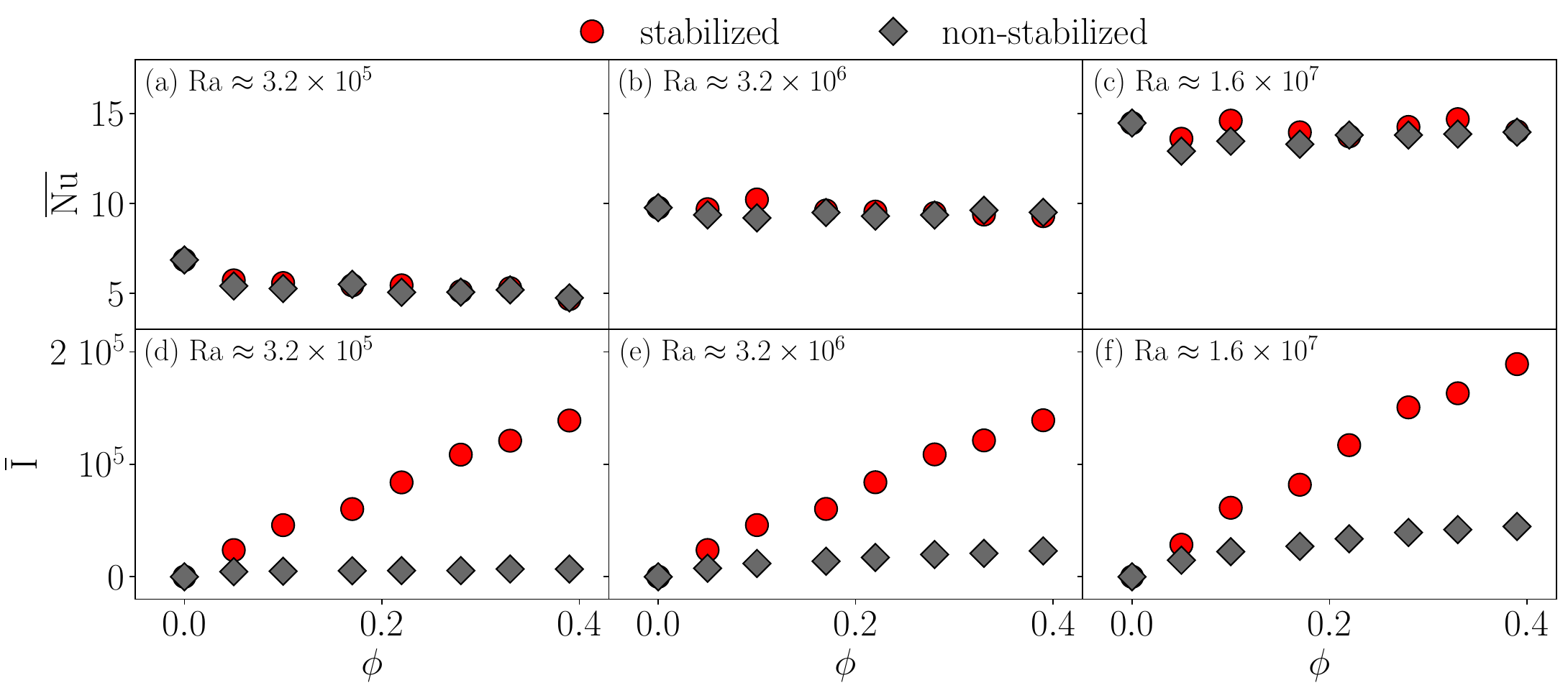}
    \caption{Steady-state averages in time of the Nusselt number $\overline{\Nu}$ (panels (a)-(c)), and the interface indicator $\bar{\I}$ (panels (d)-(f)) as a function of the volume fraction $\phi$, for different values of the Rayleigh number $\Ra$. Data for both stabilized (red circles) and non-stabilized (grey diamonds) liquid-liquid dispersions are shown.}
    \label{fig:averages_vs_phi}
\end{figure*}
We restrict our analysis to volume fractions $\phi < 0.5$, due to the impossibility to pack droplets beyond this value of concentration in the non-stabilized case; for $\phi>0.5$ the dispersion would unavoidably undergo a phase inversion with droplet (of initially continuous phase) concentration $1-\phi$, for any forcing (see Fig.~\ref{fig:comparison_phi_0.4_0.6}). In other words, non-stabilized dispersions can only exist with the minority phase being dispersed and under a non-zero forcing (otherwise
full phase separation would occur~\cite{Perlekar14}).
Stabilized dispersions, instead, can be prepared across the whole concentration range $0<\phi<1$, being stable against phase separation and prone to phase inversion only if the forcing is large enough to overcome the barrier imposed by the disjoining pressure~\cite{PelusiSM21,PelusiSM23,PelusiPRE25}. 
We collect data at varying the control parameters $\phi$ and $\Ra$. Fig.~\ref{fig:snapshots_Ra_phi} shows a comparison of density map snapshots, in the steady-state, for different combinations of the ($\phi$, $\Ra$) pair. As expected, this figure confirms that the presence or absence of an interfacial stabilization mechanism crucially affects the morphology of the dispersions: the stabilized dispersion mostly consists of many, essentially round, small droplets, whereas the non-stabilized dispersion develops domains that are, on average, bigger and more anisotropic. This qualitative comparison raises important questions about how the nature of dispersion influences both heat transfer performance and morphological organization. To answer them, we investigate the time evolution of the heat transfer, encoded in the Nusselt number $\Nu$ (cfr. Eq.~\eqref{eq:Nusselt}), and the amount of interface, encoded in the interface indicator $\I$ (cfr. Eq.~\eqref{eq:II}). Results are reported in Fig.~\ref{fig:signals_vs_Ra} for different combinations of the pair ($\phi$, $\Ra$). For fixed $\Ra$ and nature of dispersion, the time evolutions of $\Nu$ for different $\phi$ (top panels) look similar, despite heat flux fluctuations appearing more pronounced in the more diluted case. This scenario is representative of a Newtonian homogeneous fluid while approaching turbulence~\cite{Heslot87,PelusiPRE25}. However, when comparing stabilized and non-stabilized systems at fixed pair ($\phi$, $\Ra$), we observe similar time evolutions of $\Nu$, a behavior which is also reflected in kinetic energy signals (data not shown). Despite the time evolutions of $\Nu$ look qualitatively similar (see also Fig.~\ref{fig:averages_vs_phi}), the scenario changes when we analyze the dispersion morphology via the interface indicator $\I$ at fixed pair ($\phi$, $\Ra$). Indeed, the time evolutions of $\I$ (bottom panels) highlight the distinct physical mechanisms that lead to the steady-state configuration. On the one hand, the stabilized liquid-liquid dispersion does not sensibly vary its morphology unless the buoyancy amplitude is large enough and the system enters a breakup-dominated regime~\cite{PelusiPRE25}. On the other hand, the non-stabilized liquid-liquid dispersion is inherently prone to coalescence at any value of $\Ra$, leading to a progressive reduction of the total interfacial area over time, until the steady-state is reached. To summarize, in the dilute regime, a variation in $\phi$ or the nature of dispersion does not appear to impact the global heat transfer properties of the system; however, the amount of interface evolves according to different physical mechanisms, leading to distinct steady-state configurations.\\
\begin{figure*}[t!]
    \centering    \includegraphics[width=.8\textwidth]{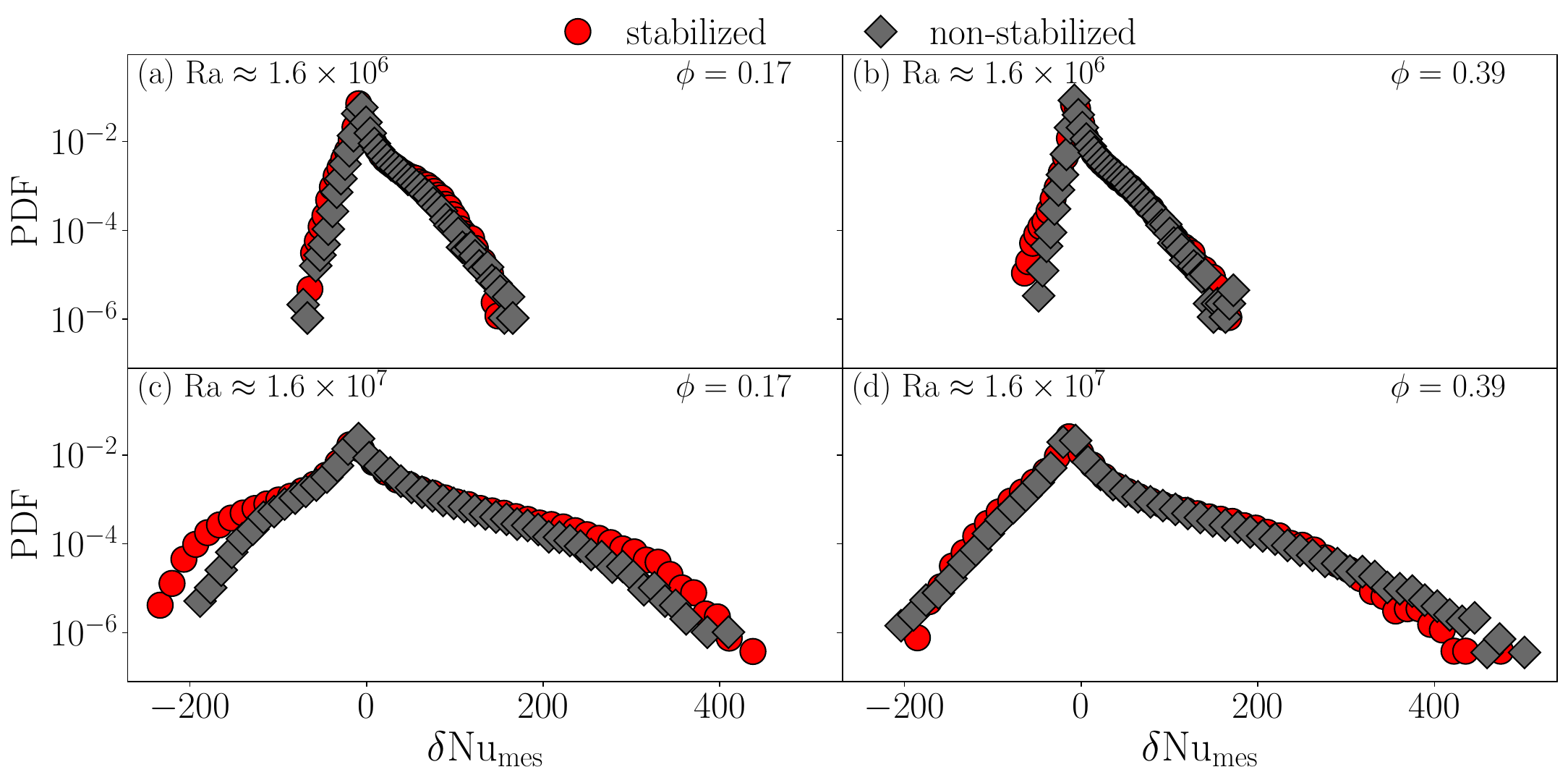}
    \caption{Probability distribution functions (PDFs) of the fluctuations $\delta \Nu_{\mathrm{mes}}$ of the Nusselt number at mesoscales $\Nu_{\mathrm{mes}}$ (cfr. Eq.~\eqref{eq:Nu_loc}) for different values of the volume fraction (different columns) and different values of the Rayleigh number $\Ra$ (different rows).  Data for both stabilized (red circles) and non-stabilized (grey diamonds) liquid-liquid dispersions are shown.}
    \label{fig:PDF_nu_local}
\end{figure*}
\begin{figure*}[t!]
    \centering
    \includegraphics[width=.8\textwidth]{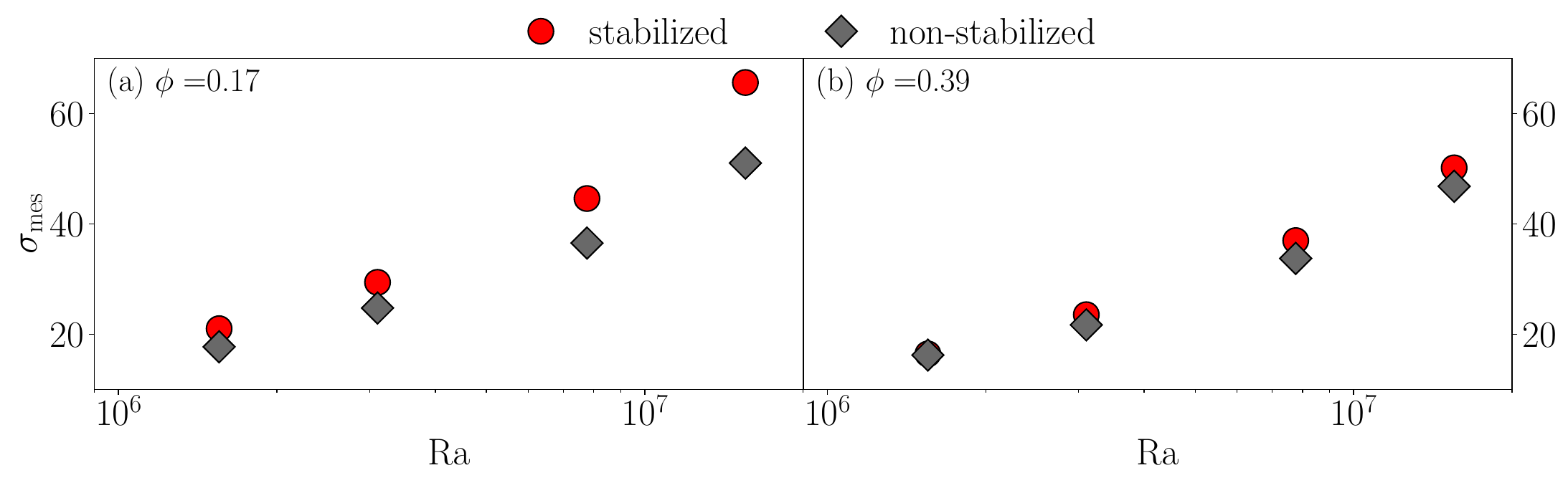}
    \caption{Standard deviation $\sigma_\mathrm{mes}$ of the Nusselt number at mesoscales $\Nu_{\mathrm{mes}}$ (see text for details) as a function of the Rayleigh number $\Ra$ for different values of the volume fraction $\phi$. Data for both stabilized (red circles) and non-stabilized (grey diamonds) liquid-liquid dispersions are shown.
    }\label{fig:sigma_vs_Ra}
\end{figure*}
We next investigate the time average over the steady state of the relevant observables, for various combinations of the ($\phi$, $\Ra$) pair. In Fig.~\ref{fig:averages_vs_phi}, we report $\overline{\Nu}$ and $\overline{\I}$ as a function of $\phi$, for different values of $\Ra$. First, we notice that, in the range of $\Ra$ explored, $\overline{\Nu}$ (which grows with $\Ra$, as expected) is essentially independent of the volume fraction and the nature of dispersion [Fig.~\ref{fig:averages_vs_phi}(a)-(c)]. A naive interpretation of this behavior might suggest that, for low to moderate volume fractions ($\phi < 0.5$), where non-Newtonian effects are absent or negligible, an increase of $\phi$ leads to a higher effective viscosity of the system and, consequently, to a monotonic decrease in the average heat flux. However, this simplistic view neglects the crucial role played by the presence of finite-size droplets in modulating heat transfer. Considering emulsions in the range $\phi < 0.5$, as previously observed in Ref.~\cite{PelusiSM21}, the presence of finite-sized droplets introduces mesoscopic interfacial structures that actively interfere with the formation of large-scale thermal plumes and inhibit the convective transport. This inhibition becomes more pronounced at low values of $\Ra$, i.e., just above the transition from conduction to convection, where the reduction in $\overline{\Nu}$ with increasing $\phi$ (within $10\%$) is evident but not as steep as one would expect based on viscosity arguments only~\cite{PelusiSM21}. However, as $\Ra$ increases (see Fig.~\ref{fig:averages_vs_phi}), this reduction in heat transfer is progressively mitigated, indicating a nontrivial interplay between $\phi$ and $\Ra$. In contrast, recent results by Bilondi et al.~\cite{Brandt24} focused on non-stabilized liquid-liquid dispersions at sufficiently high values of $\Ra$, and report an enhancement of $\overline{\Nu}$ with increasing $\phi$, attributed to the amplification of small-scale mixing induced by interfacial stresses. Putting everything together, one would then expect the existence of a non-monotonic transition in the heat transfer behavior when the system is diluted ($\phi < 0.5$), which is governed by both droplet concentration and buoyancy forcing: when $\Ra$ is low, $\overline{\Nu}$ decreases upon increasing $\phi$; then, for intermediate values of $\Ra$, the heat transfer experiences almost no variation as a function of $\phi$; finally, if the buoyancy is large enough, then is starts to increase. In this context, the heat transfer budget approach introduced in Ref.~\cite{Brandt24} — which decomposes convective and diffusive contributions from both phases — provides a valuable conceptual tool for interpreting the competing mechanisms governing this transition. In agreement with this view, our results 
suggest that droplet-scale interfacial dynamics contribute to a reorganization of the energy transport process, subtly modifying how heat is distributed between boundary layers and bulk flow as both $\phi$ and $\Ra$ vary.
Although the interfacial stabilization does not have a significant impact on the mean heat transfer, it affects the system morphology.
\begin{figure*}[t!]
    \centering
    \includegraphics[width=.8\textwidth]{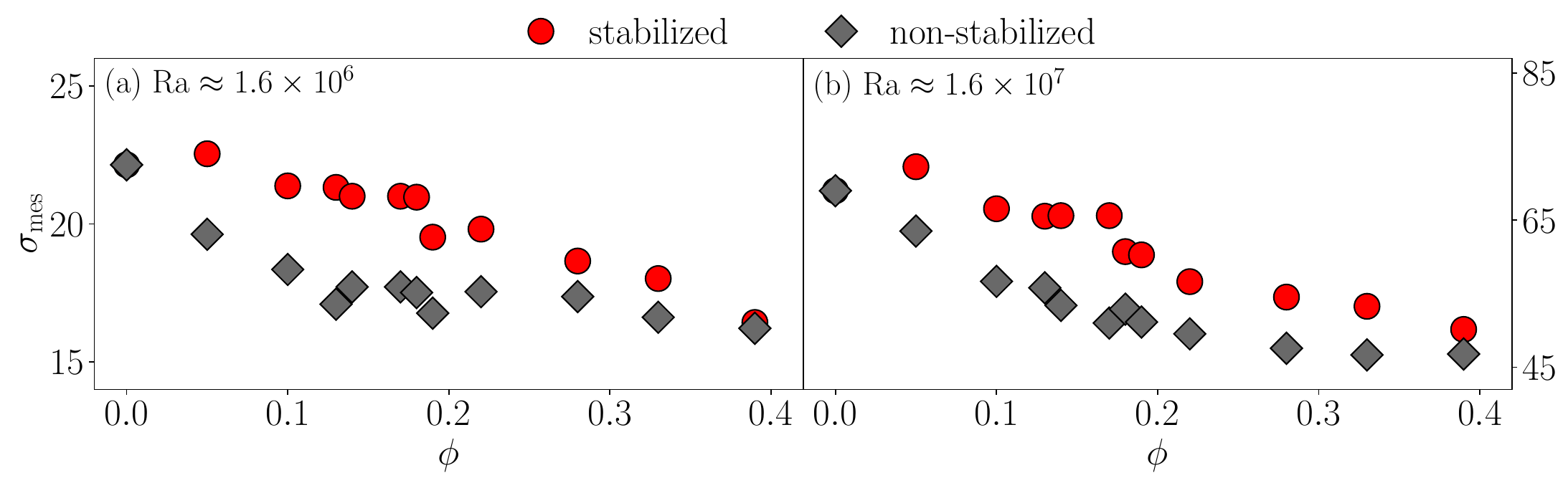}
    \caption{Standard deviation $\sigma_\mathrm{mes}$ of the Nusselt number at mesoscales $\Nu_{\mathrm{mes}}$ (cfr. Eq.~\eqref{eq:Nu_loc}) as a function of the volume fraction $\phi$ for different values of the Rayleigh number $\Ra$. Data for both stabilized (red circles) and non-stabilized (grey diamonds) liquid-liquid dispersions are shown.}\label{fig:sigma_vs_phi}
\end{figure*}
\begin{figure*}[t!]
    \centering
    \includegraphics[width=.8\textwidth]{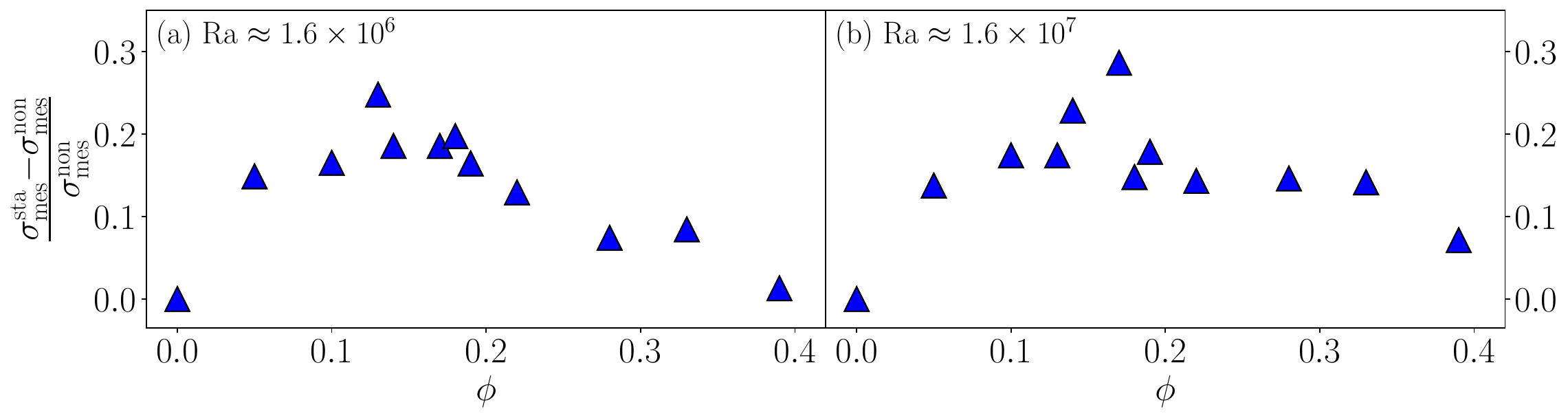}    \caption{Normalized difference between standard deviation of the Nusselt number at mesoscales $\Nu_{\mathrm{mes}}$ (cfr. Eq.~\eqref{eq:Nu_loc}) for non-stabilized ($\sigma^{\mathrm{non}}_\mathrm{mes}$) and stabilized ($\sigma^{\mathrm{sta}}_\mathrm{mes}$) liquid-liquid dispersions as a function of the volume fraction $\phi$ and for different values of the Rayleigh number $\Ra$.} \label{fig:sigma_vs_phi_normalized}
\end{figure*}
In Fig.~\ref{fig:averages_vs_phi}(d)-(f), this effect is quantified via the analysis of $\overline{\I}$ as a function of $\phi$ for different values of $\Ra$. In the case of the stabilized liquid-liquid dispersion, $\overline{\I}$ exhibits a clear increasing trend with $\phi$ across all values of $\Ra$. By contrast, in the non-stabilized case, this dependence is much weaker and becomes noticeable only at the highest $\Ra$. This difference directly stems from the inhibition of coalescence in the stabilized dispersion, which preserves a more structured and interface-rich morphology. It is therefore reasonable to expect that such pronounced morphological differences may leave a trace in the heat transfer dynamics, not necessarily in global observables, but rather through small-scale signatures associated with interfacial processes. Bearing these observations in mind, we move beyond macroscopic observables and examine heat transfer at the mesoscale, i.e., at the characteristic scale of the droplet size. Unlike previous studies on emulsions~\cite{PelusiSM21,PelusiSM23,PelusiPRL24}, where droplet-based observables could be defined thanks to the fact that droplets could be identified and followed for a reasonable amount of time during their evolution with a lagrangian tool of analysis~\cite{TLBfind22}, here a different approach is needed, since in non-stabilized liquid-liquid dispersions approaching droplets inevitably coalesce, and hence they can't be followed for long times. Thus, we focus on spatial fluctuations of the Nusselt number at mesoscales, defined as $\delta \Nu_{\text{mes}} = \Nu_{\text{mes}} - \langle \Nu_{\text{mes}} \rangle$, where $\langle \cdot \rangle$ denotes a combined spatial and temporal average. The mesoscale observable $\Nu_{\text{mes}}$ is defined as
\begin{equation}\label{eq:Nu_loc}
    \Nu_{\text{mes}}({\bm X}_k)= \int\int_{\mathcal{A}_{\Delta}({\bm X}_k)} \Nu({\bm x},t)d{\bm x}
\end{equation}
i.e., as the spatial average of the local Nusselt number, defined in Eq.~\eqref{eq:Nu_space-time}, over a square region $\mathcal{A}_{\Delta}({\bm X}_k) = [X_k-\Delta/2,X_k+\Delta/2]\times[Z_k-\Delta/2,Z_k+\Delta/2]$ centered at position ${\bm X}_k$, with $k$ running over the number of cells [$k=1,\dots,(L/\Delta)^2$]. The cell size is fixed to $\Delta = L/28 = H/28 \approx 70$. It is chosen to represent the typical mesoscale of the system: small enough to resolve local features, but large enough to accomodate -- on average -- the size of a single droplet (based on the initialization diameter $d$) surrounded by a reasonable amount of continuous phase. The corresponding probability distribution functions (PDFs) of $\delta \Nu_{\text{mes}}$ for both the stabilized and non-stabilized liquid-liquid dispersions are reported in Fig.~\ref{fig:PDF_nu_local} for some combinations of the ($\phi$, $\Ra$) pair. The $\Nu$ PDFs display the typical shape skewed towards positive values~\cite{Gasteuil07,PelusiSM21,PelusiSM23}; for lower values of $\Ra$, no relevant mismatch is observed between the two kinds of dispersion, regardless of the value of $\phi$. In contrast, at higher values of $\Ra$, the PDF of the stabilized dispersion develops more pronounced tails in the more dilute case ($\phi=0.17$), but, curiously, this mismatch is again suppressed as $\phi$ increases.
\begin{figure*}[t!]
    \centering
    \includegraphics[width=.8\textwidth]{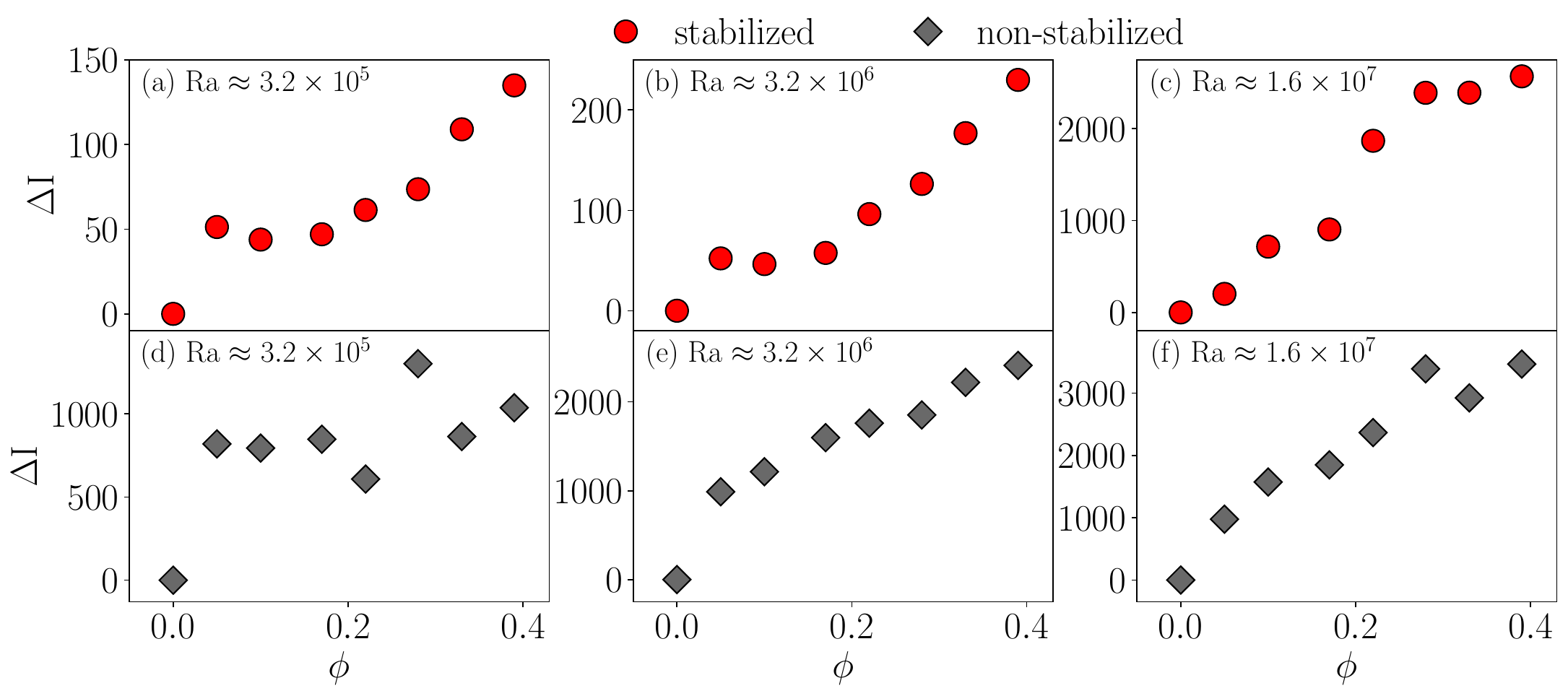}    \caption{Fluctuations of the interface indicator $\Delta \I$ (see text for details), estimated from signals reported in Fig.~\ref{fig:signals_vs_Ra}, as a function of the volume fraction $\phi$ for different values of the Rayleigh number $\Ra$. Data for both stabilized (red circles) and non-stabilized (grey diamonds) liquid-liquid dispersions are shown.} \label{fig:deltainterface_vs_phi}
\end{figure*}
This non-trivial behaviour prompted us to perform a more systematic study in the $(\phi, \Ra)$ space, by investigating the standard deviation of the PDFs of $\delta \Nu_{\text{mes}}$, namely:
\begin{equation}\label{eq:sigmaloc}
 \sigma_{\text{mes}} = \sqrt{\langle \left(\Nu_{\text{mes}} - 
 \langle \Nu_{\text{mes}} \rangle\right)^2\rangle}.
\end{equation}
The fluctuations $\sigma_{\text{mes}}$ are reported in Fig.~\ref{fig:sigma_vs_Ra} as 
a function of $\Ra$ for two representative values of $\phi$. In both cases, $\sigma_{\text{mes}}$ grows with $\Ra$, reflecting the expected rise in turbulence intensity. However, in the more diluted regime [see Fig.~\ref{fig:sigma_vs_Ra}(a)], the fluctuations in the stabilized dispersion are systematically larger than those in the non-stabilized case, with the gap widening as $\Ra$ increases. A naive interpretation might link this trend to the total amount of interface area $\overline{\I}$, which grows more steeply with $\Ra$ in the stabilized system [see Fig.~\ref{fig:averages_vs_phi}(d)–(f)]. Yet this argument does not hold at higher volume fractions [Fig.~\ref{fig:sigma_vs_Ra}(b)], where the difference in $\sigma_{\text{mes}}$ between the two systems becomes negligible, despite a further increase in interface area with $\phi$. In Fig.~\ref{fig:sigma_vs_phi} we also show the dependence of $\sigma_{\text{mes}}$ on $\phi$ for $\Ra=1.6 \times 10^6$ [panel (a)] and $\Ra=1.6 \times 10^7$ [panel(b)]. We observe that, as the volume fraction $\phi$ increases, a clear separation gradually develops between the stabilized and non-stabilized dispersions, the latter consistently displaying higher values of $\sigma_{\text{mes}}$, which reflect in stronger mesoscale heat flux fluctuations. As $\phi$ increases and the dispersion becomes more concentrated, the gap eventually closes again, indicating that at high volume fractions the distinction between the two dispersions weakens. However, this gap does not arise from an increase in fluctuations in the stabilized dispersion. Indeed, two distinct trends can be identified. In the non-stabilized case, $\sigma_{\text{mes}}$ initially decreases quite steadily and then remains constant for $\phi \gtrsim 0.2$, since droplet coalescence and restructuring suppress thermal transport. In the stabilized dispersion, $\sigma_{\text{mes}}$ slowly decreases with $\phi$. Overall, this entails a non-monotonic behaviour of the relative mismatch, as evidenced in Fig.~\ref{fig:sigma_vs_phi_normalized}, with a maximum around $0.1 < \phi < 0.2$ which weakly depends on $\Ra$.\\
A qualitative interpretation of the observed phenomenology relies on the link between heat flux fluctuations, quantified by $\sigma_{\text{mes}}$ in Eq.~\eqref{eq:sigmaloc}, and the small-scale structure of the velocity field. In the absence of stabilization against coalescence, the non-stabilized liquid-liquid dispersion will develop domains of the dispersed phase across a range of length scales. These structures can interact with turbulent fluctuations, effectively ``dissipating" kinetic energy through interface rearrangements such as coalescence and breakup. In contrast, in stabilized dispersion, where droplets remain relatively small and resist merging, the interface dynamics becomes increasingly decoupled from the turbulent flow, particularly at low volume fractions, making the fluctuations $\sigma_{\text{mes}}$ independent from $\phi$. As $\Ra$ increases and the system becomes more concentrated, the frequency of droplet collisions rises, thereby enhancing the rate of coalescence. The latter promotes the formation of larger droplets, which are more prone to breakup, thus reactivating a dynamical balance between breakup and coalescence. This renewed interfacial activity draws energy from small-scale velocity fluctuations, which are consequently attenuated, resulting in a decrease of $\sigma_{\text{mes}}$ with increasing $\phi$. These facts are also supported by the study of fluctuations in the interface indicator, $\Delta \I$, which serves as a proxy for the number of breakup and coalescence events. Fig.~\ref{fig:deltainterface_vs_phi} shows data of $\Delta \I$ as a function of $\phi$ for different values of $\Ra$. In non-stabilized liquid-liquid dispersions, the fluctuations $\Delta \I$ continuously increase with $\phi$; contrariwise, for the stabilized case, $\Delta \I$ is almost pinned up to $\phi \approx 0.2$ for any value of $\Ra$, and then increases, a behaviour that is symptomatic of the resistance of the stabilized dispersion to morphological variations induced by the increase of collision/coalescence rate.
\section{Conclusions}\label{sec:conclusions}
\begin{figure*}[t!]
    \centering    \includegraphics[width=1.\textwidth]{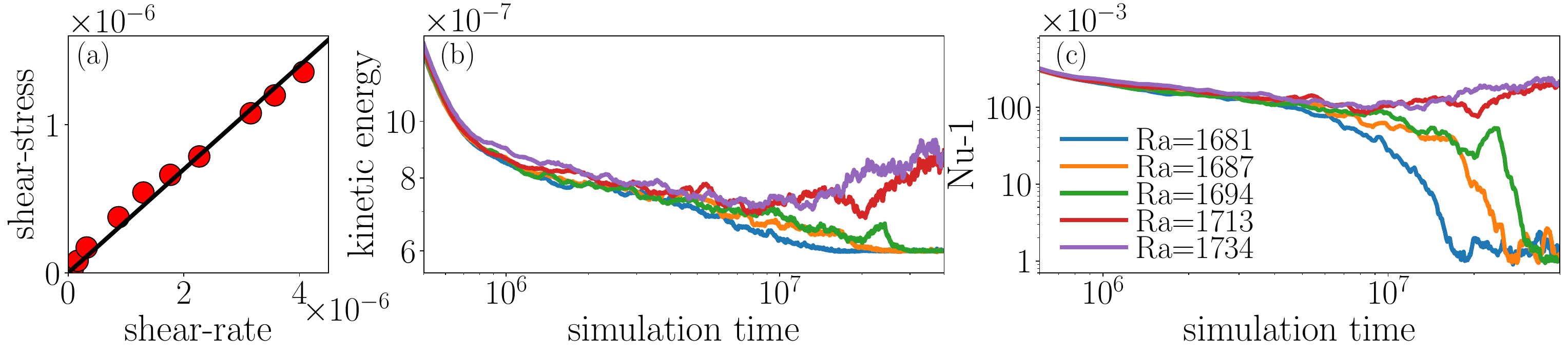}
    \caption{Panel (a): Rheological curve showing shear-stress vs shear-rate for the stabilized emulsion at $\phi=0.17$. Space-averaged kinetic energy (panel (b)) and Nusselt number (panel (c)) as a function of time for various $\Ra$. Notice that both the kinetic energy and the values of $\Nu-1$ tend to zero for $\Ra < \Ra_c$ and to a positive, finite, plateauing value for $\Ra > \Ra_c$, indicating the transition to a convective state.}\label{fig:benchmark}
\end{figure*}
We conducted a numerical investigation of the Rayleigh-Bénard thermal convection in dispersions of liquid droplets in another liquid phase by comparing two different systems: a stabilized liquid-liquid dispersion (i.e., a proper emulsion), where a mechanism of interfacial stabilization against droplet coalescence is introduced -- mimicking the effect of surfactants via a positive disjoining pressure~\cite{ravera2021} -- and a non-stabilized liquid-liquid dispersion, where the stabilization mechanism is not present. By systematically varying the droplet volume fraction $\phi$ and the Rayleigh number $\Ra$, we explored both the system morphology and the heat transfer properties. Despite significant differences in the morphological evolution of the two systems, particularly in the interface indicator $\I$, we find that the global heat transfer, quantified via the time-averaged Nusselt number $\overline{\Nu}$, remains remarkably similar between stabilized and non-stabilized liquid-liquid dispersions across the explored parameter space. This result underscores the robustness of the macroscopic thermal transport properties, even in the presence of dramatically different interfacial physics, at least in the range of $\Ra$ explored. Our analysis further reveals that small-scale features are much more sensitive to the nature of the dispersion. In particular, the fluctuations of the mesoscale Nusselt number, $\delta \Nu_{\mathrm{mes}}$, are systematically larger in stabilized dispersions, where droplet collisions and interface-mediated interactions sustain enhanced velocity and temperature gradients at the mesoscale. Conversely, in non-stabilized dispersions, coalescence leads to the suppression of small-scale fluctuations, especially at values of $\phi > 0.2$. Interestingly, the difference in mesoscale heat flux fluctuations between the two systems exhibits a non-monotonic behavior as a function of $\phi$: it peaks around $0.1 < \phi < 0.2$ and diminishes at higher concentrations. This behavior can be understood in terms of the coupling between velocity fluctuations and interface dynamics at small scales, which differs between the two systems. These observations enrich the scenario described in our previous findings on the transition between breakup- and coalescence-dominated regimes in proper emulsions~\cite{PelusiPRE25}, and are consistent with recent results at higher $\Ra$ highlighting fluctuation-driven heat transfer enhancement in non-stabilized liquid-liquid dispersions~\cite{Brandt24}.\\
Overall, our results underscore the necessity of considering macroscopic observables in conjunction with other interfacial properties when analyzing the role of different interfacial properties in thermally convective multiphase systems. Despite global observables may appear insensitive to interfacial details, interfacial physics plays a critical role in shaping the small-scale structure and energy redistribution mechanisms of the flow. Future work will focus on further characterizing stabilized dispersions by studying systems with different strengths of the disjoining pressure in the interfacial interactions. This work will help clarify better how the phenomenology observed for stabilized dispersions sets in starting from a non-stabilized one. Moreover, we plan to investigate the role of the initial configuration, particularly in stabilized emulsions where variations in $H/d$ at fixed $\phi$ may influence the ensuing dynamics. Additionally, extending the present analysis to three-dimensional systems could provide new insights into how dimensionality affects droplet dynamics and heat transport in thermally driven multiphase flows with complex interfacial properties.

\appendix

\section{Method validation}
To validate our model in the framework of thermally driven flows, we measured the critical Rayleigh number $\Ra_c$ associated with the transition from conductive to convective states for the stabilized system at $\phi=0.17$, $\Gamma=2$. The value of $\Gamma$ is required in order to compare with exact linear stability results~\cite{Chandrasekhar61}. The value of $\phi$ was chosen because an accurate determination of $\Ra_c$ requires computing the effective viscosity from the rheological curve (see Fig.~\ref{fig:benchmark}(a)), which -- as discussed in Sec.~\ref{sec:setup} -- becomes shear-rate dependent at larger concentrations. 
This approach of assuming an effective medium for the two-fluid system is made possible if the spatial distribution of droplets is homogeneous, in a statistical sense. In the non-stabilized case, which is intrinsically prone to coalescence of droplets and full phase separation, such statistical homogeneity is lost, therefore a similar comparison with exact results under the assumption of an effective medium with an effective viscosity cannot be performed.
For this benchmark, we monitored the evolution of the kinetic energy and the advective contribution to the Nusselt number (i.e., $\Nu-1$) as functions of time for different values of $\Ra$. The results, reported in 
Fig.~\ref{fig:benchmark}(b)-(c), show that our model correctly captures the onset of convection, yielding a critical Rayleigh number in the range $1694 < \Ra_c < 1713$, in excellent agreement with the theoretical prediction $\Ra_c \approx 1707$~\cite{Chandrasekhar61}.

\section*{Acknowledgements}
The authors thank Chao Sun for fruitful discussions. FP, MS, and MB acknowledge the support of the National Center for HPC, Big Data and Quantum Computing, Project CN\_00000013 – CUP E83C22003230001 and CUP B93C22000620006, Mission 4 Component 2 Investment 1.4, funded by the European Union – NextGenerationEU. Support from INFN/FIELDTURB project is also acknowledged.

\bibliography{francesca}

\end{document}